\documentclass[12pts,twocolumn]{revtex4-1} 
\usepackage{subfigure}
\usepackage{amsmath} 
\usepackage{amsthm,amsfonts}
\usepackage{graphicx,color}
\usepackage[normalem]{ulem}

\bibpunct{[}{]}{,}{n}{}{}

\begin{document}


\title{Boundary element method and simulated annealing algorithm applied to \\ electrical impedance tomography image reconstruction \\ \textit{\small{{(M\'etodo dos elementos de contorno e algoritmo de recozimento simulado aplicados na \\ reconstru\c{c}\~ao de imagem da tomografia de imped\^ancia el\'etrica)}}}}

\author{Olavo H. Menin}
\affiliation{Instituto Federal de Educa\c c\~ao, Ci\^encia e Tecnologia de S\~ao Paulo,\\
R. Am\'erico Ambr\'osio, 269, 14169-263, Sert\~aozinho, SP, Brazil.}
\email{olavohmenin.ifsp@gmail.com} 

\author{Vanessa Rolnik}
\affiliation{Faculdade de Filosofia  Ci\^encias e Letras de Ribeir\~ao Preto, Universidade de S\~ao Paulo, 
\\Av. Bandeirantes, 3900, 14040-901, Ribeir\~ao Preto, SP, Brazil.}

\author{Alexandre S. Martinez}
\affiliation{Faculdade de Filosofia  Ci\^encias e Letras de Ribeir\~ao Preto, Universidade de S\~ao Paulo, 
\\Av. Bandeirantes, 3900, 14040-901, Ribeir\~ao Preto, SP, Brazil. \\ National Institute of Science and Technology in Complex Systems, Brazil.}

\date{\today}

\begin{abstract}
Physics has played a fundamental role in medicine sciences, specially in imaging diagnostic. Currently, image reconstruction techniques are already taught in Physics courses and there is a growing interest in new potential applications. The aim of this paper is to introduce to students the electrical impedance tomography, a promising technique in medical imaging. We consider a numerical example which consists in finding the position and size of a non-conductive region inside a conductive wire. We review the electric impedance tomography inverse problem modeled by the minimization of an error functional. To solve the boundary value problem that arises in the direct problem, we use the boundary element method. The simulated annealing algorithm is chosen as the optimization method. Numerical tests show the technique is accurate to retrieve the non-conductive inclusion.
\bigskip

A f\'isica tem tido um papel fundamental nas ci\^encias m\'edicas, especialmente em diagn\'osticos por imagem. Atualmente, as t\'ecnicas de reconstru\c{c}\~ao de imagem j\'a s\~ao ensinadas nos curso de F\'isica e existe um crescente interesse em poss\'iveis novas aplica\c{c}\~oes.  O objetivo deste trabalho \'e apresentar aos alunos a tomografia de imped\^ancia el\'etrica, uma promissora t\'ecnica de imageamento em medicina. Para isso,  consideramos um exemplo num\'erico que consiste em encontrar a posi\c{c}\~ao e o tamanho de uma regi\~ao n\~ao condutora no interior de um fio condutor. N\'os revisamos o problema inverso da tomografia de imped\^ancia el\'etrica modelado pela minimiza\c{c}\~ao de um funcional de erro. Para resolver o problema de valor de contorno que surge no problema direto, n\'os usamos o m\'etodo dos elementos de contorno. O algoritmo de recozimento simulado foi escolhido como m\'etodo de otimiza\c{c}\~ao. Testes num\'ericos mostram que a t\'ecnica \'e precisa para encontrar a inclus\~ao n\~ao condutora. \\
\noindent{\bf Palavras-chave:} tomografia de imped\^ancia el\'etrica; m\'etodo dos elementos de contorno; algoritmo de recozimento simulado; problema inverso; otimiza\c{c}\~ao.

\end{abstract}

\maketitle


\section{Introduction} \label{sec_introduction}

\begin{figure} [!h]
\centering
\includegraphics[scale = 0.35]{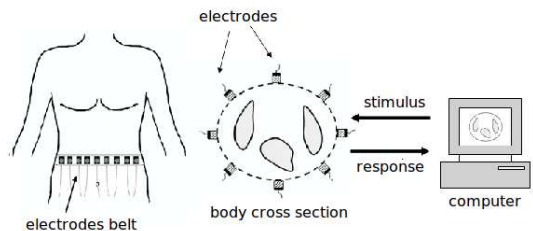}
\caption{\small{Layout of the electrical impedance tomography. An electrodes belt is attached on the body surface. A voltage source applies a stimulus on the electrodes and the corresponding response is measured and supplied to a computer. A specific software reconstructs the interior body image.}}
\label{eit_layout}
\end{figure}

Electrical impedance tomography (EIT) is a technique to obtain the inner image of an object exploring its electrical properties. It consists to apply an electric voltage (or electric current) stimulus through electrodes positioned on the object external surface. The corresponding response (electric current or voltage) is measured by the same electrodes. The obtained data are supplied to a computer, which has an algorithm to reconstruct the conductivity distribution inside the object. This distribution can be interpreted as the interior object image. A scheme of a medical EIT is depicted in Fig.~\ref{eit_layout}.
 
The first occurrence of imaging the human body in terms of its electrical properties occurred in 1978 by Henderson and Webster \cite{Henderson1978}. After this study, several research groups have emerged providing significant developments in EIT technique. Current applications of EIT in medical area are the monitoring of pulmonary functions, breast cancer detection, blood flow monitoring inside the heart and imaging of human brain functions. Also, in engineering, EIT applications includes the monitoring of two-phase flow in a pipe, detection of corrosion or cracks inside metallic objects and retrieval a pipeline route in the subsoil \cite{Lionheart2005, Abascal2008, Osella2001}.

Although the EIT technique presents attractive features, such as low cost, portability and robustness, it is not widely used yet due to the difficulty to obtain a good image resolution. This difficulty arises because the EIT image reconstruction is an inverse-problem and therefore, it is intrinsically ill-posed \cite{Borcea2002}. The Hadamard criteria, (i) existence, (ii) uniqueness and (iii) continuous data dependence on the solution are not simultaneously  guaranteed. The first criterion is satisfied,  since the physical body being imaged certainly  has a actual conductivity distribution. With respect to the second one, it has been shown for the Dirichlet-to-Neumann map there is a unique solution for the conductivity distribution inside the domain \cite{Nachman1996}. However, the third criterion has been the hardest one to be overcome. 
To deal with it, the researchers have tried to find a good and suitable regularization function or alternative approaches. For example, it is possible to restrain the search-space to speed up the convergence to the actual solution and avoid nonphysical solutions \cite{Menin2011}.
  
A typical way to solve the reconstruction EIT problem is to treat it as an optimization problem. In this case, one must minimize an error functional which is a value expressing the discrepancy between two different models of the same problem, one experimentally obtained and the other one numerically calculated. The conductivity distribution that yields the error function global minimum corresponds to the sought image. 

Due to the ill-conditioned nature of the inverse EIT problem, the optimization surface  presents irregularities such as multiple local minima and almost flat regions. Hence, one must adopt an efficient optimization procedure to handle such topological features and we have chosen the Simulated Annealing (SA) \cite{Aarts1988, Goldberg1989} . This is a stochastic method in which the main feature is the possibility of eventually accept a solution that increases the objective function. This allows the method to detrap from local minima basins or flat regions and reach the global minimum \cite{Martins2011,Kim2005}.

To carry out the iterative searching, one must solve the direct problem with different conductivity distributions. It corresponds to a boundary value problem (BVP), which models the stimulus/response process. Frequently, this BVP does not have analytical solution, requiring the use of numerical methods, such as the Finite Element Method (FEM) and the Finite Difference Method (FDM). Here, the Boundary Element Method (BEM) was chosen. This is well suited to EIT problem, since it requires only the discretization of the domain boundary and the solution on the boundary is calculated firstly without calculating the values of the electric potential inside the domain \cite{Rolnik2010,Duraiswami1998}. Moreover, BEM offers important advantages such as great flexibility for arbitrary geometries and boundary shape and easiness of implementation \cite{Brebbia1992, Ang2007}.

Our goal in this paper is to present to students the EIT technique using BEM and SA algorithm. This study was motivated by the difficulty in teaching the physical and mathematical methods/processes related to image reconstruction. Some studies have been carried in experimental approach, such as the assembly proposed by Mylott \textit{et al.} \cite{Mylott2011}  to study the computed tomography. Considering the students' growing interest in computational physics in the last years, we have proposed a numerical example to introduce the EIT technique. It consists in retrieving a non-conductive region inside a conductive wire. In engineering, this non-conductive region may represent, for example, a manufacturing defect or a normal wear in the wire which could compromise its use in an apparatus. In medicine, it may model an air bubble inside an artery hampering the blood flow.  

\begin{figure} [h!]
\centering
\includegraphics[scale=0.37]{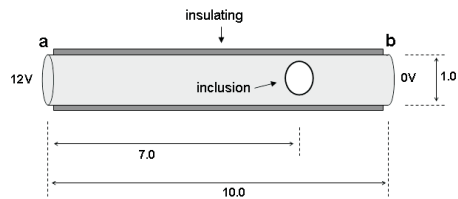}
\caption{\small{Scheme of the numerical example: cylindrical conductive wire with length $10.0$ and diameter $1.0$ with a non-conductive region of radius $0.3$ located $7.0$ units far from extremity $a$.}}
\label{wire3D}
\end{figure}

Our numerical example considers a conductive cylindrical wire of length $10.0$ and diameter $1.0$ (arbitrary units). The non-conductive region is a sphere of radius $0.3$ with center located along the cylinder axis at $7.0$ units far from one of its extremities, as shown in Fig.~\ref{wire3D}. The challenge is summarized as a two variables optimization problem, the center $x$-coordinate and radius $R$ of the non-conductive region. Due to the domain symmetry, we have taken a longitudinal section of the wire as a two-dimensional domain and placed the left bottom corner at the origin of a Cartesian system $x0y$. Also, the stimulus/response process consists in applying an electric voltage $V_{ab}=12\,\mbox{V}$ between the wire extremities $a$ and $b$ ($V_a=12\,\mbox{V}$ and $V_b=0\,\mbox{V}$) and measuring the current flux on the same extremities. Moreover, the lateral area of the wire is considered electrically insulated, such that the current flux vanishes.  

In Sec.~\ref{sec_statement}, we review the EIT mathematical modeling based on the functional approach, the Boundary Element Method and the Simulated Annealing algorithm. In Sec.~\ref{sec_tests}, we show the numerical tests and the results. Finally, in Sec.~\ref{sec_discussion}, we discuss the results and conclude.


\section{Problem statement and methods} \label{sec_statement}

Here, we present the mathematical and numerical basis of EIT problem, specially focused on our numerical example.

\subsection{EIT mathematical formulation} \label{subsec_eit}

The mathematical modeling of the EIT problem is obtained from the Maxwell's equations. Since the stimulus processing is made at low frequency, the inductive effects can be ignored. Also, considering the domain fully ohmic, we can neglect the capacitive effects. The electric field is $\vec{E} = -\vec{\nabla} \phi$, where $\phi$ is the electrical potential and $\vec{\nabla}$ is the gradient operator. The current density is $\vec{J} = - \sigma\vec{E} = -\sigma\vec{\nabla}\phi$, being $\sigma$ the electrical conductivity. Considering no internal current sources, $\vec{\nabla}\cdot\vec{J} = 0$ and the partial differential equation (PDE),

\begin{equation}
\vec{\nabla} \cdot \left( - \sigma \vec{\nabla}\phi \right)=0\qquad\mbox{in}\quad \Omega,
\label{pde}
\end{equation}

\noindent governs the electrical potential $\phi$ inside the domain $\Omega$. To simulate the stimulus/response process, we have two boundary conditions given by $\phi = \overline{\phi}$ and $-\sigma \dfrac{\partial \phi}{\partial n} = \overline{J}$, where $\overline{\phi}$ is the voltage profile and $\overline{J}$ is the normal component of the electric current flux, both on the boundary $\partial\Omega$.

The numerical example considers a two-phase medium problem. One of them (the spherical non-conductive region) has null-conductivity and the other one (the rest of the wire) has constant and uniform non-vanishing conductivity. In this case, the interior of the non-conductive inclusion is interpreted as the exterior of the domain and the conductivity distribution $\sigma$ corresponds  to the position and size of the non-conductive region. Also, Eq.~(\ref{pde}) is simplified to the Laplace equation

\begin{equation}
\nabla^2 \phi = \dfrac{\partial^2 \phi}{\partial x^2}+\dfrac{\partial^2\phi}{\partial y^2} = 0, \qquad\mbox{for}\quad (x,y) \in \Omega.
\label{eq_laplace}
\end{equation}

Considering unitary the conductivity of the non null phases ($\sigma = 1$), the boundary conditions become

\begin{equation}
\phi(x,y) = \overline{\phi}(x,y), \qquad\mbox{for}\quad (x,y) \in \partial \Omega,
\label{bc_dirichlet}
\end{equation}

\begin{equation}
-\dfrac{\partial \phi(x,y)}{\partial n} = \overline{J}(x,y), \qquad\mbox{for}\quad (x,y) \in \partial \Omega.
\label{bc_neumann}
\end{equation}

The stimulus-response can be mathematically modeled through the Dirichlet-to-Neumann map, in which a known voltage profile $\overline{\phi}$ is imposed on the boundary and the electric current fluxes $\overline{J}$ are calculated on the same boundary. Otherwise, if a known electric current flux profile $\overline{J}$ is imposed and the voltage $\overline{\phi}$ is calculated, the model is called Neumann-to-Dirichlet map. Although in practice the EIT problem is formulated as a Dirichlet-to-Neumann map, mathematically, it represents a mixed problem since it is imposed null current flux for the internal boundary, i.e., Eq.~(\ref{bc_neumann}) vanishes.

The \textit{direct problem} consists in solving the BVP given by Eqs.~(\ref{eq_laplace}) and (\ref{bc_dirichlet}), considering a knowing conductivity distribution $\sigma$ to obtain the current flux from Eq.~(\ref{bc_neumann}). However, in EIT image reconstruction, $\sigma$ is unknown and we have an \textit{inverse problem}. One approach to solve it and to obtain $\sigma$ is to construct and minimize an error functional (objective function) that compares the measurements obtained from two different models, the \textit{actual model} and the \textit{numerical model}.

For the Dirichlet-Neumann map, the \textit{actual model} corresponds to a collection of measurements, $\textbf{J}_{actual}$, which are the current fluxes on the electrodes obtained from experimental assembly. This model contains the unknown actual conductivity distribution, $\sigma_{actual}$, which must be found. The \textit{numerical model} considers a known prospective conductivity distribution, $\sigma_{prosp}$, to solve the \textit{direct problem} to obtain $\textbf{J}_{num}$, which corresponds to the numerical current fluxes on the same electrodes. The error functional must be defined as a discrepancy measure between $\textbf{J}_{actual}$ and $\textbf{J}_{num}$. Here, we have defined the error functional $e(\sigma_{prosp})$ as the mean square error between the experimental and numerical current fluxes,

\begin{equation}
e\,(\sigma_{prosp}) = \frac{1}{n}\sum\limits_{i=1}^{n} |J_{actual}^{(i)} - J_{num}^{(i)}| ^2,
\label{func_error}
\end{equation}

\noindent where $J_{actual}^{(i)}$ and $J_{num}^{(i)}$ are, respectively, the actual and numerical fluxes in the electrode $i$ and $n$ is the number of electrodes.

The idea of the optimization process is to perform an iterative search to find the prospective conductivity distribution that minimizes Eq.~(\ref{func_error}). In each iteration, a prospective solution, $\sigma_{prosp}$, is generated and supplied to Eqs.~(\ref{eq_laplace}) and (\ref{bc_dirichlet}) to yield a  numerical flux $\textbf{J}_{num}$. This flux is compared with the actual one $\textbf{J}_{actual}$ to calculate the error functional, from Eq.~(\ref{func_error}). This process is repeated so that $e\,(\sigma_{prosp})\approx 0$ and, consequently, $\sigma_{prosp} \approx \sigma_{actual}$ within an acceptable error level.

\subsection{Boundary element method}\label{subsec_bem}

The essence of BEM is to convert the field equation to integral equations on the domain boundary through the reciprocal relation,  

\begin{equation}
\int\limits_{\partial\Omega}\left(\phi_2\frac{\partial \phi_1}{\partial n} - \phi_1\frac{\partial \phi_2}{\partial n}\right)ds(x,y) = 0,
\label{reciprocal_relation}
\end{equation}

\noindent where $\phi_1$ and $\phi_2$ are two general solutions of Eq.~(\ref{eq_laplace}), and through the fundamental solution,

\begin{equation}
\Phi(x,y;\xi,\eta) = \frac{1}{4\pi}\ln[(x-\xi)^2 + (y-\eta)^2],
\label{fund_solution}
\end{equation}

\noindent which is a particular solution of Eq.~(\ref{eq_laplace}). From Eqs.~(\ref{reciprocal_relation}) and (\ref{fund_solution}), it is possible to obtain two integral equations, one for the points $(\xi,\eta)$ inside the domain $\Omega$ and another for the points $(\xi,\eta)$ on the boundary $\partial\Omega$. In the discretized form, these integral equations are

\begin{equation}
\phi(\xi,\eta)=\sum\limits_{k=1}^{N_T}\left[\overline{\phi}^{(k)}F_2^{(k)}(\xi,\eta) - \overline{J}^{(k)} F_1^{(k)}(\xi,\eta)\right],
\label{sol_discret_1}
\end{equation}

\noindent for $(\xi,\eta) \in \Omega$, and

\begin{equation}
\frac{1}{2}\overline{\phi}(\xi,\eta)=\sum\limits_{k=1}^{N_T}\left[\overline{\phi}^{(k)} F_2^{(k)}(\xi,\eta) - \overline{J}^{(k)} F_1^{(k)}(\xi,\eta)\right],
\label{sol_discret_2}
\end{equation}

\noindent for $(\xi,\eta) \in \partial\Omega$, where

\begin{eqnarray}
F_1^{(k)}(\xi,\eta) &=&\int\limits_{\partial\Omega^{(k)}}\Phi(x,y;\xi,\eta)ds(x,y), \\
F_2^{(k)}(\xi,\eta) &=& \int\limits_{\partial\Omega^{(k)}}\frac{\partial}{\partial n}\left[\Phi(x,y;\xi,\eta)\right]ds(x,y)
\end{eqnarray}

\bigskip
The external boundary is discretized in $N_{ext}$ straight elements, $\partial\Omega^{(k)}, k=1,2,...,N_{ext}$, and the internal one in $N_{int}$ elements, $\partial\Omega^{(k)}, k=N_{ext}+1,N_{ext}+2,...,N_{T}$. The coordinates $(x^{(k)},y^{(k)})$ of the elements extremities are supplied to the program  and, by convention, the elements are numbered following the counter clockwise direction for the external boundary and clockwise for the internal one. This guarantees the normal unitary vector $\vec{n}$ always points to the outside of the domain. Also, for each boundary (external and internal), the final extremity of the last element is made to coincide with the initial extremity of the first one.  In this case, the domain and the inclusion are approximated by polygons. 

Equation~(\ref{sol_discret_2}) produces a linear system of $N_T$ equations with $N_T$ unknowns, either $\overline{\phi}$ or $\overline{J}$. Solving this system provides the pair $(\overline{\phi},\overline{J})$ for all boundary elements. Hence, the direct problem solution of EIT is completed. If the solution $\phi$ at some internal point is required, it can be calculated through Eq.~(\ref{sol_discret_1}).

\subsection{Simulated annealing algorithm} \label{subsec_sa}

The simulated annealing algorithm is a stochastic optimization method that has its origins in the Metropolis acceptance criterion when two system configurations are compared \cite{Metropolis1953}. It is based on the probability to find the system with energy $E$, given by the Boltzmann weight, $e^{-E/k_B T}$, where $T$ is the temperature and $k_B$ is the Boltzmann constant. 

In 1983, Kirkpatrick \textit{et al.} \cite{Kirkpatrick1983} solved the salesman problem using the Metropolis criterion adding an important differential. They adopted a cooling schedule for the temperature to control the search stochasticity. In the thermodynamic framework, if the temperature of a liquid material is slowly cooled down, the atoms arrange themselves to form a structure (a perfect crystal), which has the lowest internal energy state. However, if the cooling process is not sufficiently  slow, the final  structure  is not perfect and the internal energy is not the lowest one. 

The analogy with an optimization process is made considering the objective function $f (\vec{x})$ as the energy $E$ of the system, the solutions $\vec{x}_i$ as the system configurations and the temperature $T$ becomes a control parameter of the process. The optimization  process is done iteratively starting from an initial solution $\vec{x}_0$ and temperature $T_0$. At each iteration, $k$ searches are performed. Each new solution, $\vec{x}_{i+1}$ ($i=0,1,2,...,k$), is generated through a pre-defined visitation distribution and compared with the current one, $\vec{x}_i$, to be, or not, accepted according to the Metropolis criterion. If $f(\vec{x}_{i+1}) \leq f(\vec{x}_i)$, the new solution is accepted and if $f(\vec{x}_{i+1}) > f(\vec{x}_i)$, the new solution is accepted with probability

\begin{equation}
p(\Delta f, T)=e^{-\Delta f/T},
\label{prob_SA}
\end{equation}

\noindent where $\Delta f = f(\vec{x}_{i+1})-f(\vec{x}_i)$ and $k_B = 1$. After the last solution $\vec{x}_k$ is evaluated, the temperature is decreased through a cooling schedule and the search process restarts. The last solution accepted at the previous temperature is taken as the initial solution to the current one.

The most common ways to generate a new solution $x_{i+1}$ from the previous one $x_{i}$ are through uniform or Gaussian deviates. 
Here, we have chosen the Gaussian one in the following form

\begin{equation}
x_{i+1}=x_i+\left(1+\dfrac{T_t}{T_0}\right)\lambda\, \eta,
\label{distrib_gauss}
\end{equation}

\noindent where $T_0$ is the initial temperature, $T_t$ is the temperature at the iteration $t$, $\lambda$ is a parameter that depends on the search space range and $\eta$ is a random number with normal distribution $N(0,1)$. 

According to Eqs.~(\ref{prob_SA}) and (\ref{distrib_gauss}), at high temperatures, solutions can be generated relatively far from the previous one, characterizing a global search. Also, there is high probability to accept a new solution that increases the objective function. As the temperature decreases, the search becomes essentially local and the acceptance of an uphill error solution becomes less likely to occur. Hence, the temperature $T$ plays a fundamental role. The choice of a suitable cooling schedule is essential to ensure a good performance of the optimization process, i.e., to reach the global minimum of the objective function with as few steps as possible. The most commons cooling schedules in the literature are the logarithmic and the geometric ones. We have chosen the last one in which the temperature $T$ decreases with the iteration $t$ according to 

\begin{equation}
T_t = \alpha^t\,T_0,
\label{geometric_cooling}
\end{equation}

\noindent where $T_0$ is the initial temperature and $\alpha \in (0,1)$ is the cooling rate.


\section{Numerical tests and results}\label{sec_tests}

To solve the numerical example proposed in Sec.~\ref{sec_introduction}, we developed in MATLAB$^{\copyright}$ language a program that includes the BEM and the SA algorithm routines. 

To construct the error functional, Eq. (\ref{func_error}), one must confront two models of the same problem, the actual and the numerical ones. The \textit{actual model} is simulated numerically, since it is not the goal of this paper to explore the experimental part of the problem. 
Hence, the actual current flux, $\textbf{J}_{actual}$, is obtained solving numerically the Eq.~(\ref{eq_laplace}) through BEM with the boundary conditions, $\overline{\phi}\,(0,y) = 12\,V$ and $\overline{\phi}\,(10,y) = 0\,V$, for $y \in [0,1]$, and $\overline{J}\,(x,0) = \overline{J}\,(x,1) = 0$, for $x \in [0,10]$. 
Also, it was adopted $\overline{J}\,(x,y) = 0$ for the points $(x,y)$ on the boundary of the non-conductive spherical region. 
The \textit{numerical model} is defined as the actual one. 
However, its non-conductive region, called \textit{prospective inclusion}, has center $x$-coordinate ($x_p$) and radius ($R_p$) variables in order to carry out the searching process. 

For both models, the external boundary discretization was made using $220$ elements, being $100$ elements for each side, top and bottom, and $10$ elements for each side, left and right. 
The internal inclusion was discretized in $80$ elements, totalizing $300$ boundary elements. 

\subsection{Test 1 - Error functional behavior}

\begin{figure} [h!]
\centering
\includegraphics[scale=0.365]{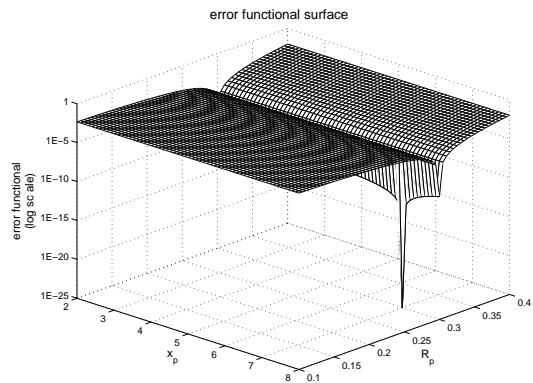}
\caption{\small{Error functional surface in logarithm scale. It presents almost flat regions and a channel that contains a prominent global minimum.}}
\label{error_surf}
\end{figure}

The first test explores the behavior of the error functional surface. The variables $x_p$ and $R_p$ were systematically modified over the search space $[2.0,8.0] \times [0.1,0.4]$ in a regular mesh of $61\times61 = 3721$ points. The error functional is calculated for each position $x_p$ and radius $R_p$ generating the error functional surface, as shown in the plot of Fig.~\ref{error_surf} with error functional axis in logarithm scale. It is possible to see almost flat regions and a narrow channel that contains a prominent global minimum. These characteristics require a powerful optimization method to reach the global minimum. 

\subsection{Test 2 - Applying the Simulated annealing}

In this test, we evaluated the performance of the developed program to retrieve the position and radius of the non-conductive region. Since SA algorithm is stochastic, its convergence to the actual solution is not always guaranteed. Therefore, it is necessary to run the program many times to assess its performance. Also, to make a more careful analysis, we have divided the \textbf{Test 2} in three parts.

In the first part, \textbf{Test 2.a}, the intent was to find only the center $x$-coordinate of the non-conductive region. We set the prospective inclusion radius equal to the actual one ($R_p = 0.3$). The initial solution of $x_p$ was generated randomly in the range $[1.0,9.0]$ and, to generate a new solution $x_{p}^{(k+1)}$ from the previous one $x_{p}^{(k)}$ by Eq.~(\ref{distrib_gauss}), it was chosen  $\lambda = 0.8$. In the second one, \textbf{Test 2.b}, the program searched only the non-conductive region radius. In this case, we set the prospective inclusion center $x$-coordinate equal to the actual one ($x_p = 7.0$). The initial solution of $R_p$  was generated randomly, but in the range $[0.05,0.45]$. The new solution $R_{p}^{(k+1)}$ from the previous one $R_{p}^{(k)}$ by Eq.~(\ref{distrib_gauss}) was generated considering $\lambda = 0.04$. For both, \textbf{Test 2.a} and \textbf{Test 2.b}, the program was executed considering $1000$ iterations, with $k = 1$, and we set $\alpha = 0.95$ and $T_o = 1000$.

Finally, the third part, \textbf{Test 2.c}, considers the search for both, position and radius. In this case, we joined the two programs used in the previous parts in a single one. More specifically, in each iteration (at each temperature) the program carried out two separated searching, one for the position and another for the radius. Due the difficulty to carry out the optimization searching with two variables, we adopted $2000$ iterations, with $k = 1$, and $\alpha =  0.97$. As the previous parts, we set $T_0 = 1000$,  $\lambda = 0.8$, to generate a new $x_p$, and $\lambda = 0.04$, to generate a new $R_p$. Again, the initial solutions of $x_p$ and $R_p$ were generated randomly in the same ranges of the previous parts.

\begin{figure}[h!]
\centering
\includegraphics[scale=0.38]{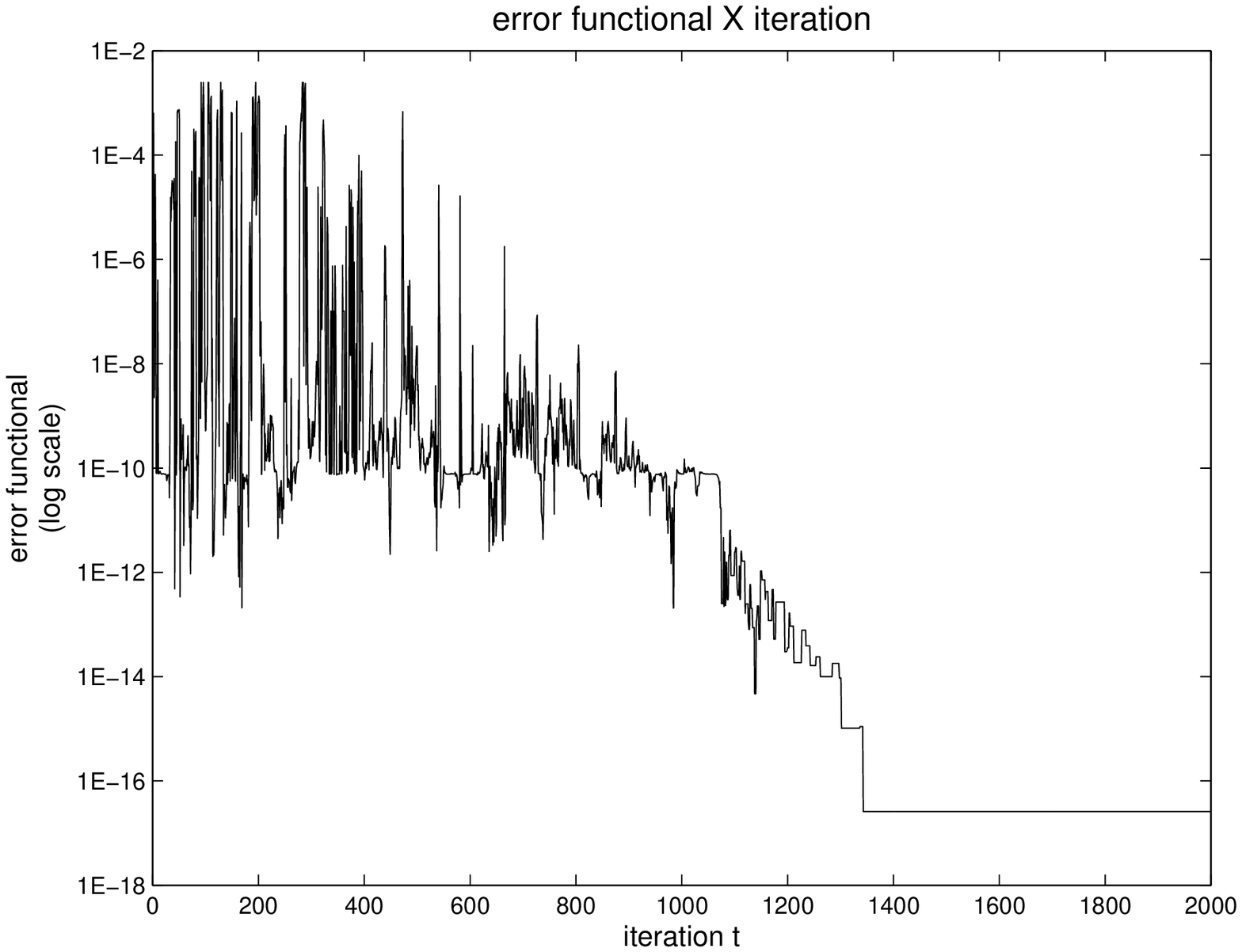}\\
\small{(a)}\\
\includegraphics[scale=0.38]{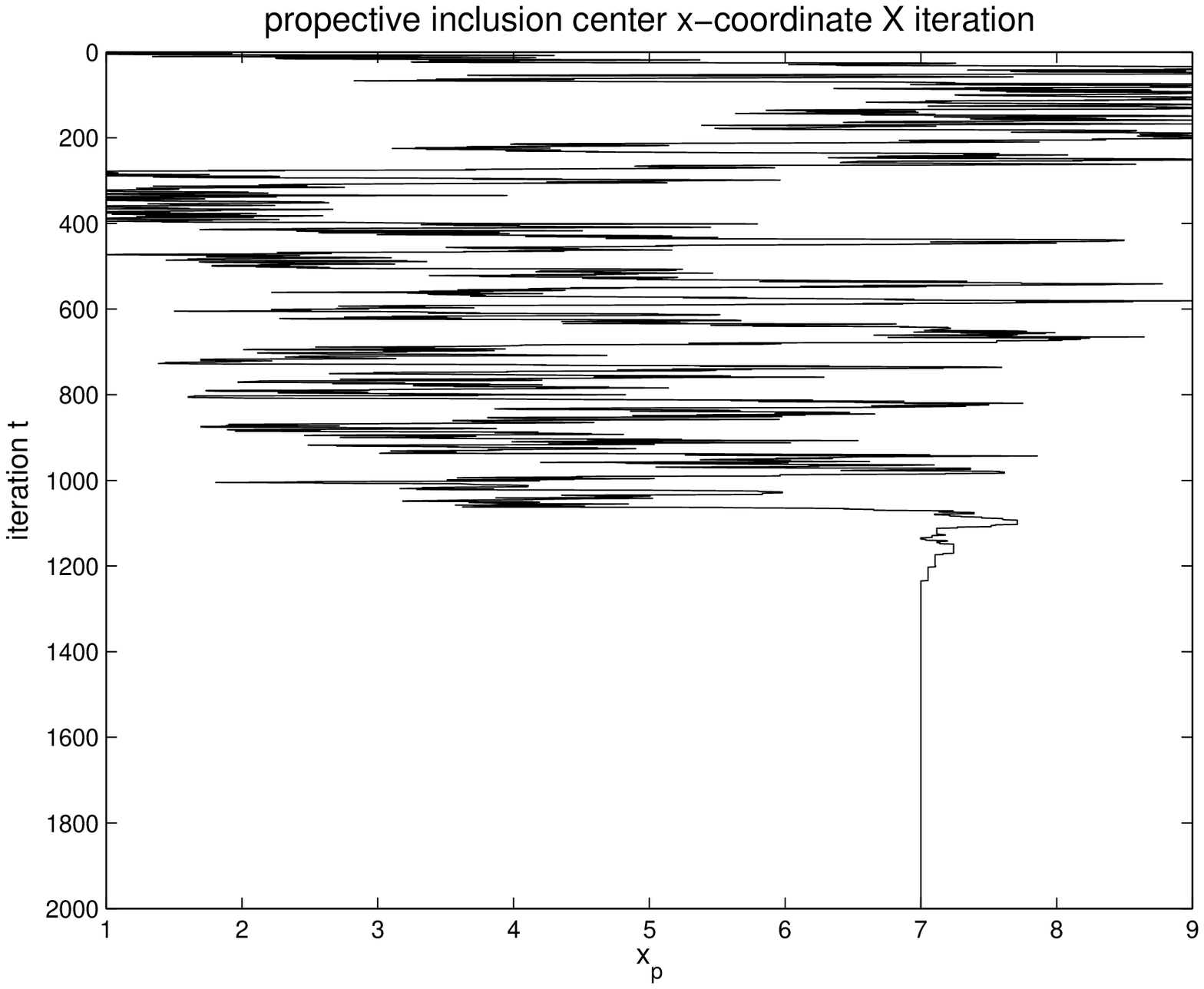}\\
\small{(b)}\\
\includegraphics[scale=0.38]{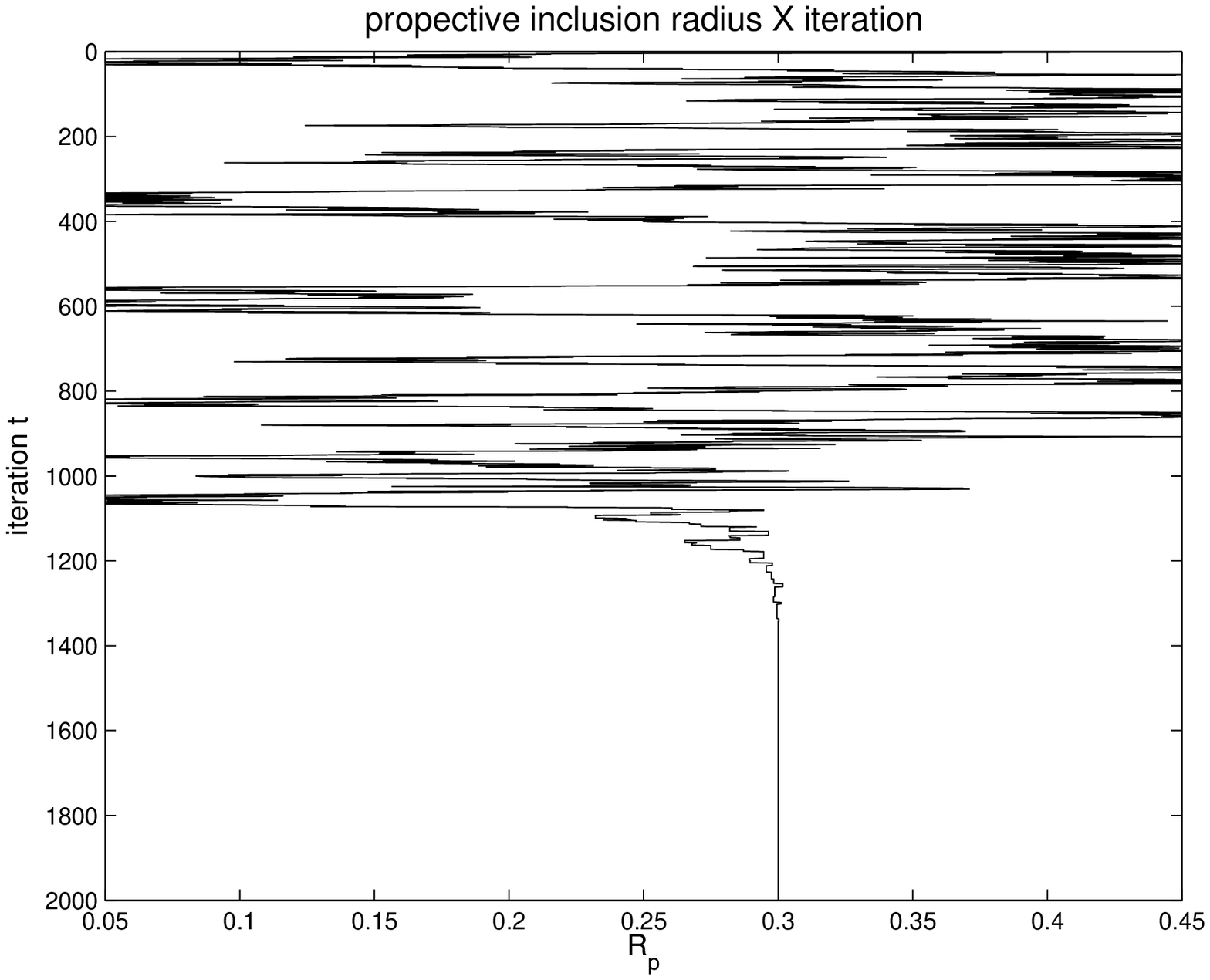}\\
\small{(c)}\\

\caption{\small{Evolution with the iteration for the best running of \textbf{Test 2.c}  of the (a) error functional (b) prospective center $x$-coordinate and (c) prospective radius.}} 
\label{graph_bestrunning}
\end{figure}

For each part of the \textbf{Test 2}, the program was executed $50$ times and the final solution $x_p$ and/or $R_p$ found in each one was stored. Table~\ref{table} shows the minimum, maximum, mean and standard deviation of the final solutions considering all of the runnings for each part. For the best running of the \textbf{Test 2.c}, the solution found was $x_p = 7.0005$ and $R_p = 0.3000$, with error functional of $2.6\times10^{-17}$. For the best running, the evolution of the error functional, the prospective center $x$-coordinate and the prospective radius during the optimization process are shown in Fig.~(\ref{graph_bestrunning}).

\begin{table}[ht]
\centering
\caption{\small{Minimum, maximum, mean and standard deviation of the final solutions $x_p$ and/or $R_p$ for the three parts of the \textbf{Test 2}, considering $50$ runnings.}}
\label{table}
\begin{tabular} {l|c|c|c|c}

         &\multicolumn{1}{|c|}{\small{Test 2.a}} & \multicolumn{1}{|c|}{\small{Test 2.b}} & \multicolumn{2}{|c}{\small{Test 2.c}} \\ 
\hline
         & \small{$x_p$}    & \small{$R_p$}     &\small{$x_p$}     & \small{$R_p$}    \\ 
\hline
\small{min}      & \small{$6.9925$}  & \small{$0.2995$}   & \small{$6.8388$}   & \small{$0.2445$} \\ 
\hline
\small{max}      & \small{$7.0081$}  & \small{$0.3005$}   &   \small{$7.5467$} & \small{$0.3163$}  \\ 
\hline
\small{mean}     & \small{$7.0000$}  &   \small{$0.3001$} &  \small{$7.1203$}  & \small{$0.2876$}  \\ 
\hline
\small{std}      & \small{$0.0032$}  &   \small{$0.0002$} &  \small{$0.2053$}  & \small{$0.0214$} \\ 

\end{tabular}
\end{table}


\section{Discussion and conclusion} \label{sec_discussion}

We have presented to students the electrical impedance tomography technique and the application of retrieving the size and position of an non-conductive inclusion inside of a conductive wire using the boundary element method and simulated annealing algorithm. Numerical simulations assessed either the behavior of the error functional in the search-space and the performance of the implemented program. The results show that the problem has a difficult error surface to be optimized. Despite this, the implemented program has presented a satisfactory accuracy to find the non-conductive region. Finally, the presented algorithms and methods have a wide scope of application and should be better appreciated in topics of undergraduate and graduate physics programs.

 
\section*{Acknowledgments}

ASM acknowledges Conselho Nacional de Desenvolvimento Cient\'ifico e Tecnol\'ogico (CNPq) (305738/2010-0). VR acknowledges Funda\c{c}\~ao de Amparo \`a Pesquisa do Estado de S\~ao Paulo (FAPESP) (2008/01284-4).


%

\end{document}